\documentclass[]{spie}  

\usepackage{amsmath,amsfonts,amssymb}
\usepackage{graphicx}
\usepackage[colorlinks=true, allcolors=blue]{hyperref}
\usepackage{gensymb}

\title{Direct Classification of Type 2 Diabetes From Retinal Fundus Images in a Population-based Sample From The Maastricht Study}

\author[a]{Friso G. Heslinga}
\author[a]{Josien P.W. Pluim}
\author[c]{A.J.H.M. Houben}
\author[c]{Miranda T. Schram}
\author[c]{Ronald M.A. Henry}
\author[c]{Coen D.A. Stehouwer}
\author[c]{Marleen J. van Greevenbroek}
\author[a,b]{Tos T.J.M. Berendschot}
\author[a]{Mitko Veta}

\affil[a]{Eindhoven University of Technology, Department of Biomedical Engineering, the Netherlands}
\affil[b]{Maastricht University Medical Center+, University Eye Clinic Maastricht, the Netherlands}
\affil[c]{Maastricht University Medical Center+, Department of Internal Medicine and CARIM, the Netherlands}

\authorinfo{Send correspondence to F.G. Heslinga. E-mail: f.g.heslinga@tue.nl, Telephone: +31 (0)40 247 55 37}

\begin{document} 
\maketitle

\begin{abstract}
Type 2 Diabetes (T2D) is a chronic metabolic disorder that can lead to blindness and cardiovascular disease. Information about early stage T2D might be present in retinal fundus images, but to what extent these images can be used for a screening setting is still unknown. In this study, deep neural networks were employed to differentiate between fundus images from individuals with and without T2D. We investigated three methods to achieve high classification performance, measured by the area under the receiver operating curve (ROC-AUC). A multi-target learning approach to simultaneously output retinal biomarkers as well as T2D works best (AUC = 0.746 [$\pm$0.001]). Furthermore, the classification performance can be improved when images with high prediction uncertainty are referred to a specialist. We also show that the combination of images of the left and right eye per individual can further improve the classification performance (AUC = 0.758 [$\pm$0.003]), using a simple averaging approach. The results are promising, suggesting the feasibility of screening for T2D from retinal fundus images.
\end{abstract}


\section{INTRODUCTION}
Type 2 Diabetes mellitus (T2D) is a chronic metabolic disorder characterized by hyperglycemia, insulin resistance, and relative insulin deficiency. Late detection of T2D can lead to long-term damage, including blindness \cite{Congdon03} and cardiovascular disease \cite{Gu99}. Although T2D diagnosis based on blood-glucose measurements works well, half of the people living with diabetes worldwide were undiagnosed in 2017. \cite{Cho18} This is unfortunate, especially since major health benefits are expected from early detection and treatment. \cite{Herman15}
Non-invasive, easy-accessible screening methods could improve early detection. 

Retinal fundus imaging is widely used for the detection of diabetic retinopathy (DR), one of the complications of T2D. Over the last few years, deep learning (DL) has been proposed for automated analysis of retinal fundus images. \cite{Ting19} DR is relatively unambiguous and DL models have shown excellent detection performance. For example, Gulshan et al. \cite{Gulshan16} obtained an area under the receiver operating curve (ROC-AUC) of 0.99 for detection of referable DR.

Despite the promising results for DR detection, retinal fundus images are not used for early T2D detection, even though the vascular geometrical structures of the retina have been related to early T2D.\cite{Zhang17} The aim of this study is to investigate to what extent a deep learning model is able to distinguish T2D and non-T2D cases in retinal fundus images and to evaluate what techniques can be used to improve the classification. 

\subsection{Related work}
To the best of our knowledge, only one study has explored the value of deep learning for direct classification of T2D.\cite{Abbasi17} In addition, Poplin et al.\cite{Poplin18} used deep learning to extract cardiovascular risk factors from retinal fundus images, including a key diagnostic measure for T2D,  haemoglobin A1c (HbAIc). While high predictive performance was reported for age and sex, and some predictive information was found for smoking history and systolic blood pressure, model predictions for HbAIc levels correlated poorly with the HbAIc labels (R\textsuperscript{2} = 0.09).

Others have focused on the extraction of handcrafted features from fundus images.\cite{Romeny16,DashtBozorg16} Features such as vessel tortuosity, mean arteriolar width and venular width are considered biomarkers for T2D \cite{Zhang17}. In previous work we showed that these biomarkers can be approximated with a deep learning approach \cite{Heslinga19}. In this study we investigated the added value of these biomarkers for the training process of a deep learning model that directly classifies fundus images. 

\section{METHODS}
The color fundus images used for this research originate from The Maastricht Study\footnote{https://www.demaastrichtstudie.nl/}, an observational prospective population-based cohort study. The rationale and methodology have been described elsewhere.\cite{Schram14} Eligible for participation in The Maastricht Study were all individuals aged between 40 and 75 years and living in the southern part of the Netherlands. The study population was enriched with T2D participants for reasons of statistical power. For our study only images from individuals with T2D and normal glucose metabolism were included (8924 images from 2336 individuals in total). Other diabetes types and prediabetes individuals were excluded.

The data was divided into sets for training, validation and testing according to a 60\%/20\%/20\% split. All images of a single individual were assigned to the same set. An overview of the sets is shown in Tabel~\ref{data}. The sets comprise images of left and right eyes that are centered either on the fovea or on the optic disc. The images were resized to 1024 x 1024 pixels and channel-wise global contrast normalization was applied before further processing.

\begin{table}[ht]
\caption{Data set split details.} 
\label{data}
\begin{center}       
\begin{tabular}{|l|l|l|l|l|}
\hline
\rule[-1ex]{0pt}{3.5ex}  & Training & Validation & Test & Total \\
\hline
\rule[-1ex]{0pt}{3.5ex} \textbf{Total number of individuals}   & 1376 & 464  & 496  & 2336\\
\hline
\rule[-1ex]{0pt}{3.5ex} \textbf{age (years) [$\pm$std]}  & 60.0 [$\pm$8.5] & 59.6 [$\pm$8.1] &  60.4 [$\pm$8.2]  & 59.9 [$\pm$8.2] \\
\hline
\rule[-1ex]{0pt}{3.5ex} \textbf{sex (\% men)}  & 47.2 & 50.2 & 54.1 & 51.2\\
\hline
\rule[-1ex]{0pt}{3.5ex} \textbf{T2D individuals} & 466 (33.9\%)   & 159 (34.3\%)  & 182 (36.7\%)  & 807 (34.5\%) \\
\hline
\rule[-1ex]{0pt}{3.5ex} \textbf{Number of images} &  5222 & 1802 & 1900 & 8924\\
\hline
\end{tabular}
\end{center}
\end{table} 

All experiments were performed using DL models based on a VGG-19 architecture for which the output layer was replaced. Data-augmentation was used to expand the number of training images, encompassing translation (0 - 20 pixels), rotation (0 - 360\degree), horizontal and vertical reflection, intensity shift(0 - 20/256), color shift (0 - 30/256) and contrast shift (0 - 0.1). Inputs for the DL models are 800 x 800 pixels centered crops of the 1024 x 1024 augmented images. Models were implemented in Keras\cite{chollet2015keras} using a TensorFlow\cite{tensorflow2015-whitepaper} backend and training was done with balanced batches of 18 on 3 GPU's. Optimization of the model weights was done using Adam. Target labels are either 0 = normal glucose metabolism or 1 = T2D. The best performing model was selected based on the validation set. Final performance of the models was evaluated by the ROC-AUC on the test set

\subsection{Model setup and initialization}
First, we evaluated the effect of initialization of the model's weights for the classification of T2D images versus non-T2D images. We compared five different strategies: (1) random initialization; (2) ImageNet weights; (3) model pretrained on global retinal microvascular measurements (T2D biomarkers), including vessel caliber and vessel tortuosity\cite{Heslinga19}; (4) A multi-target learning (MTL) approach with random initialization and (5) Multi-target learning with ImageNet weights. For the T2D biomarker approach (3) we first trained a model to predict four microvascular measures as described elsewhere\cite{Heslinga19} and then replaced the output layer for the classification task.
For the multi-target approaches (4 and 5), we simultaneously predicted four T2D biomarkers and T2D status. The learning rate schedule and L\textsubscript{2}-regularization were optimized on the validation set after which all experiments were repeated three times using different random seeds to obtain a measure for standard deviation.

\subsection{Aleatoric uncertainty estimation}
In a clinical setting one can decide to refer an image for further inspection if the assessor is too uncertain about the decision. Ahyan et al. \cite{Ahyan18} showed that test-time augmentation (TTA) can be used to define a measure for the aleatoric uncertainty. We applied 30-fold TTA to the model that performed best on the validation set using the same augmentation settings as applied during training to find the posterior distribution of the T2D predictions. We used variance of the prediction distribution, \textit{var(Pred)}, as a measure for aleatoric uncertainty. Additionally, proximity of the mean of the prediction distribution to 0.5, \textit{abs(mean(Pred)-0.5)}, was evaluated as a measure of uncertainty, since this is exactly half-way the labels of healthy and T2D. We show the effect of the referral of images that the model is uncertain about, by excluding these from the results and recalculating the ROC-AUC for different referral fractions.

\subsection{Individual-level estimation}
Multiple fundus images (1 to 12) were available per individual, providing a similar number of T2D predictions. Different strategies for the aggregation of image-level predictions to individual-level predictions were evaluated for the model that performed best on the validation set: (1) mean of the soft predictions for the left and right eye; (2) maximum of the predictions for the left and right eye; (3) logistic regression and (4) Gaussian Naive Bayes. For the machine learning techniques (3 and 4) the following features were selected: Mean, variance and number of images for each of the combinations left/right eye and optic disc and fovea centered images, resulting in 12 features per individual. Average padding was used for missing values: If for one eye no optic disc centered image or fovea-centered images was available, the prediction for the opposite-centered image was used. If no image was available for one eye, the prediction for the other eye was used.

\section{RESULTS}
An overview of the results for different model setups and weight initialization is shown in Tabel~\ref{init}. If a single (non-augmented) image was used for evaluation, the ROC-AUC was found to be in the range of 0.726-0.739. When 30-fold TTA was applied, the ROC-AUC slightly increased for all strategies, with the best performance found for the MTL approach with randomly initialized weights (AUC = 0.746 [$\pm$0.001]).

\begin{table}[ht]
\caption{Model setup and initialization results.} 
\label{init}
\begin{center}       
\begin{tabular}{|l|l|l|}
\hline
\rule[-1ex]{0pt}{3.5ex} \textbf{Initialization}  & ROC-AUC [$\pm$std] & 30-fold ROC-AUC [$\pm$std] \\
\hline
\rule[-1ex]{0pt}{3.5ex} \textbf{random initialization}   & 0.726 [$\pm$0.006] & 0.729 [$\pm$0.009]  \\
\hline
\rule[-1ex]{0pt}{3.5ex} \textbf{ImageNet weights} & 0.733 [$\pm$0.003]  & 0.737 [$\pm$0.008]  \\
\hline
\rule[-1ex]{0pt}{3.5ex} \textbf{T2D biomarker weights} & 0.734 [$\pm$0.004] & 0.738 [$\pm$0.006]\\
\hline
\rule[-1ex]{0pt}{3.5ex} \textbf{MTL w. random initialization} & 0.733 [$\pm$0.010] & \textbf{0.746 [$\pm$0.001]} \\
\hline
\rule[-1ex]{0pt}{3.5ex} \textbf{MTL w. ImageNet weights} &  0.739 [$\pm$0.002] & 0.741 [$\pm$0.001] \\
\hline
\end{tabular}
\end{center}
\end{table} 

\begin{figure} [h]
\begin{center}
\begin{tabular}{c}
\includegraphics[height=8cm]{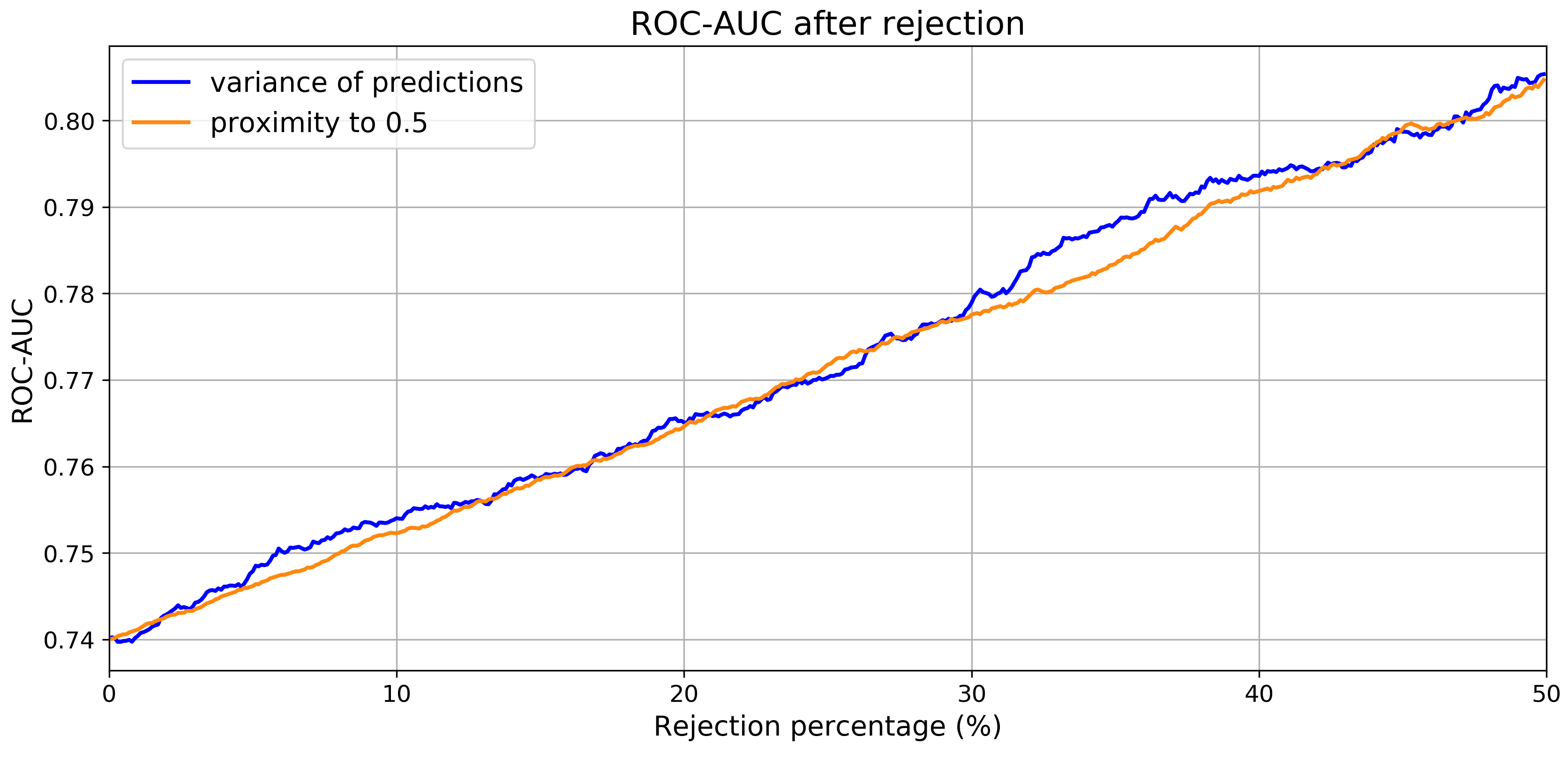}
\end{tabular}
\end{center}
\caption[example] 
{ \label{rejection} ROC-AUC after rejection of images with high prediction uncertainty}
   \end{figure}

The model that performed best on the validation set was one of the MTL models with ImageNet weights. Its performance on the test set was found to be 0.740 with 30-fold TTA. When a fraction of the images was left out for referral, based on high uncertainty of the prediction for those images, the ROC-AUC substantially increased (Figure~\ref{rejection}). For example, when 20\% of the images was excluded, the ROC-AUC increased to 0.765. Interestingly, the effect on the ROC-AUC seemed similar for both uncertainty measures. 

The combination of multiple images to obtain an individual-level prediction resulted in a higher ROC-AUC (e.g. 0.758 [$\pm$0.003] for mean of both eyes) than for single images (0.733 [$\pm$0.010]). The use of more complex classifiers did not lead to significantly better classification performance than a simple mean over the images of the left and right eye, as is shown in Tab.~\ref{subject_level}

\begin{table}[ht]
\caption{Individual-level evaluation.} 
\label{subject_level}
\begin{center}       
\begin{tabular}{|l|l|l|l|l|l|}
\hline
\rule[-1ex]{0pt}{3.5ex}  & image-level & mean of left & max of left & logistic & Gaussian Naive \\ 
& & and right eye & and right eye & regression & Bayes \\
\hline
\rule[-1ex]{0pt}{3.5ex} \textbf{ROC-AUC [$\pm$std]}   & 0.733 [$\pm$0.010] & 0.758 [$\pm$0.003] & 0.755 [$\pm$0.005] & 0.761 [$\pm$0.004] & 0.757 [$\pm$0.002] \\
\hline
\end{tabular}
\end{center}
\end{table} 

\section{CONCLUSION AND DISCUSSION}
Individuals with type 2 diabetes can be distinguished quite well from individuals with normal glucose metabolism in The Maastricht Study population using retinal fundus images and deep learning techniques. Minor benefits can be expected from optimization of the model setup and weight initialization. We found that an MTL approach with randomly initialized weights works marginally better than the other models. Classification performance improvement can be achieved with referral of the most uncertain cases and the use of multiple images per individual. This result is in line with the finding of Leibig et al.\cite{Leibig17} who leveraged prediction uncertainty to successfully refer fundus images with signs of diabetic retinopathy that were difficult to grade. This step will however lead to the referral of more false positives, which could hemper the cost-effectiveness in a screening setting.

One possibility to use retinal fundus imaging as a screening technique is the use of smartphone fundus photography.\cite{Haddock13} Future research is needed to evaluate the value of the addition of basic patient characteristics, such as sex, age and body mass index. The inclusion criteria should be extended to comprise early T2D cases (prediabetes), which were excluded for this research. Moreover, clinical validation on an external data set is needed to assess the value of the automated classification of fundus images in a general screening setting.

\acknowledgments 
This research is financially supported by the TTW Perspectief program and Philips Research. The authors have no conflicts of interests to report. This work has not been submitted for publication anywhere else. The clinical data used in the research originates from the Maastricht Study. The Maastricht Study was supported by the European Regional Development Fund via OP-Zuid, the Province of Limburg, the Dutch Ministry of Economic Affairs (grant 31O.041), Stichting De Weijerhorst (Maastricht, The Netherlands), the Pearl String Initiative Diabetes (Amsterdam, The Netherlands), CARIM School for Cardiovascular Diseases (Maastricht, The Netherlands), Stichting Annadal (Maastricht, The Netherlands), Health Foundation Limburg (Maastricht, The Netherlands) and by unrestricted grants from Janssen-Cilag B.V. (Tilburg, The Netherlands), Novo Nordisk Farma B.V. (Alphen aan den Rijn, The Netherlands), and Sanofi-Aventis
Netherlands B.V. (Gouda, The Netherlands).

\bibliography{report} 
\bibliographystyle{spiebib} 

\end{document}